\documentclass[11pt,a4paper]{article}
\pdfoutput=1
\usepackage{color}

\usepackage{jcappub}

\usepackage{epstopdf}

{\count255=\time\divide\count255 by 60 \xdef\hourmin{\number\count255}
  \multiply\count255 by-60\advance\count255 by\time
  \xdef\hourmin{\hourmin:\ifnum\count255<10 0\fi\the\count255}}

\newcommand{\gsim}{\mbox{\raisebox{-1.ex}{$\stackrel
     {\textstyle>}{\textstyle\sim}$}}}

\def\vev#1{ \left\langle #1 \right \rangle }

\newcommand{\impc}{\mathrm{Mpc}^{-1}}
\providecommand{\e}[1]{\ensuremath{\times 10^{#1}}}

\begin{document}

\title{The Knotted Sky I: Planck constraints on the primordial power spectrum}

\author[a]{Grigor Aslanyan,}
\author[a]{Layne C. Price,}
\author[b]{Kevork N. Abazajian,}
\author[a]{and Richard Easther}

\affiliation[a]{Department of Physics, University of Auckland, Private Bag 92019, Auckland, New Zealand}
\affiliation[b]{Department of Physics, University of California at Irvine,
  Irvine, CA 92697\vspace{4pt} }

\emailAdd{g.aslanyan@auckland.ac.nz}
\emailAdd{lpri691@aucklanduni.ac.nz}
\emailAdd{kevork@uci.edu}
\emailAdd{r.easther@auckland.ac.nz}

\abstract{
  Using the temperature data from \emph{Planck} we search for departures from a power-law primordial power spectrum, employing Bayesian model-selection and posterior probabilities.  We parametrize the spectrum with $n$ knots  located at arbitrary values of $\log{k}$, with both linear and cubic splines.  This formulation  recovers both slow modulations and sharp transitions in the primordial spectrum.  The power spectrum is well-fit by a featureless, power-law at wavenumbers $k>10^{-3} \, \impc$. A modulated primordial  spectrum yields a better fit relative to $\Lambda$CDM  at large scales, but there is no strong evidence for a departure from a power-law spectrum.  Moreover, using simulated maps we show that a local feature at $k \sim 10^{-3} \, \impc$ can mimic the suppression of large-scale power.  With multi-knot spectra  we see  only small changes in the posterior distributions for the other free parameters in the standard  $\Lambda$CDM universe.  Lastly, we investigate whether  the hemispherical power asymmetry is explained by independent features in the primordial power spectrum in each ecliptic hemisphere, but find no significant differences between them.
}


\maketitle

\section{Introduction}

The first cosmological data analysis by the \emph{Planck} Science Team \cite{Collaboration:2013ww} confirmed the conventional $\Lambda$CDM model of cosmology with unprecedented precision.  In particular, a scale-invariant primordial power spectrum (PPS) is  excluded at  $>5\sigma$.  Simple models of  single-field inflation generically yield an almost scale-invariant PPS, but no inflationary models are yet favoured by Bayesian evidence relative to $\Lambda$CDM  \cite{Martin:2010hh,Easther:2011yq,Collaboration:2013vu,Martin:2013nzq}. Conversely,  models with relatively complex spectra, including  oscillations or localized amplifications, are consistent with current cosmological data \cite{Ashoorioon:2006wc,Ashoorioon:2008qr,Flauger:2009ab,Achucarro:2010da,Peiris:2013opa,Easther:2013we}. 

While it is clear that the Harrison-Zel'dovich spectrum does not provide an optimal fit to the data, it does not follow that the power-law PPS is preferred over all other possible forms. Furthermore, as cosmological constraints become sensitive to increasingly delicate signals in the PPS, it is important to check whether constraints on these parameters depend  on the assumed form of the PPS. Model-independent approaches to  reconstructing the PPS have been widely studied \cite{Wang:2013vf,2013JCAP...07..031H,2013JCAP...12..035H,2012JCAP...10..050G,2010ApJ...711....1P,2010JCAP...01..016N,2009JCAP...07..011N,2009PhRvD..79d3010N,2008PhRvD..78l3002N,2008PhRvD..78b3511S,2007PhRvD..75l3502S,2006MNRAS.367.1095T,2006MNRAS.372..646L,2004PhRvD..70d3523S,2004JCAP...04..002H,Efstathiou:2003bh,2003ApJ...599....1M,2001PhRvD..63d3009H,Verde:2008er,Peiris:2009ke,Bird:2010mp,Bridges:2005br,Bridges:2006zm,Bridges:2008ta,2012JCAP...06..006V,Vazquez:2011xa,Vazquez:2013dva,dePutter:2014vd}, and the approached used here closely parallels that of Ref.~\cite{2012JCAP...06..006V}, which examines the seven year WMAP dataset.  

We revisit this problem using \emph{Planck} data, Bayesian model-selection  based on evidence (or \emph{marginalized likelihood}) ratios \cite{Bridges:2005br,Bridges:2006zm,Bridges:2008ta,2012JCAP...06..006V,Vazquez:2011xa,Vazquez:2013dva} and a  flexible specification for the PPS.  We use this formalism to test whether \emph{Planck} constraints on cosmological parameters are weakened when permit a generic PPS, rather than usual, almost--scale-invariant power-law formulation. While parameter degeneracies with the PPS could, in principle affect the posteriors on other cosmological variables (e.g., \cite{Kinney:2001js}), we find that the constraints on these parameters do not change significantly when we allow generic forms of the PPS. Secondly,  we  use this formalism to determine whether the observed large-scale hemispherical asymmetry in the two ecliptic hemispheres can be attributed to differences in the form of the PPS.  We find no difference in the structure of the power spectrum in the two hemispheres, either qualitatively or in the evidence ratios. Finally, the algorithm described here was  implemented in  \textsc{Cosmo++} \cite{Aslanyan:2013ts}, which is publicly available.

 We compare  Bayesian evidence for the non--power-law models to the evidence for the red-tilted PPS of $\Lambda$CDM. As we add more parameters to the PPS the Bayesian evidence does not change significantly, indicating the data cannot substantially distinguish between these models.  However, most of the extra knots appear in the long wavelength section of the power spectrum, with $k \lesssim 10^{-3} \, \impc$, suggesting that smaller scales are indeed well described by a power-law PPS.

\begin{figure}
\centering
\includegraphics[width=0.75\textwidth]{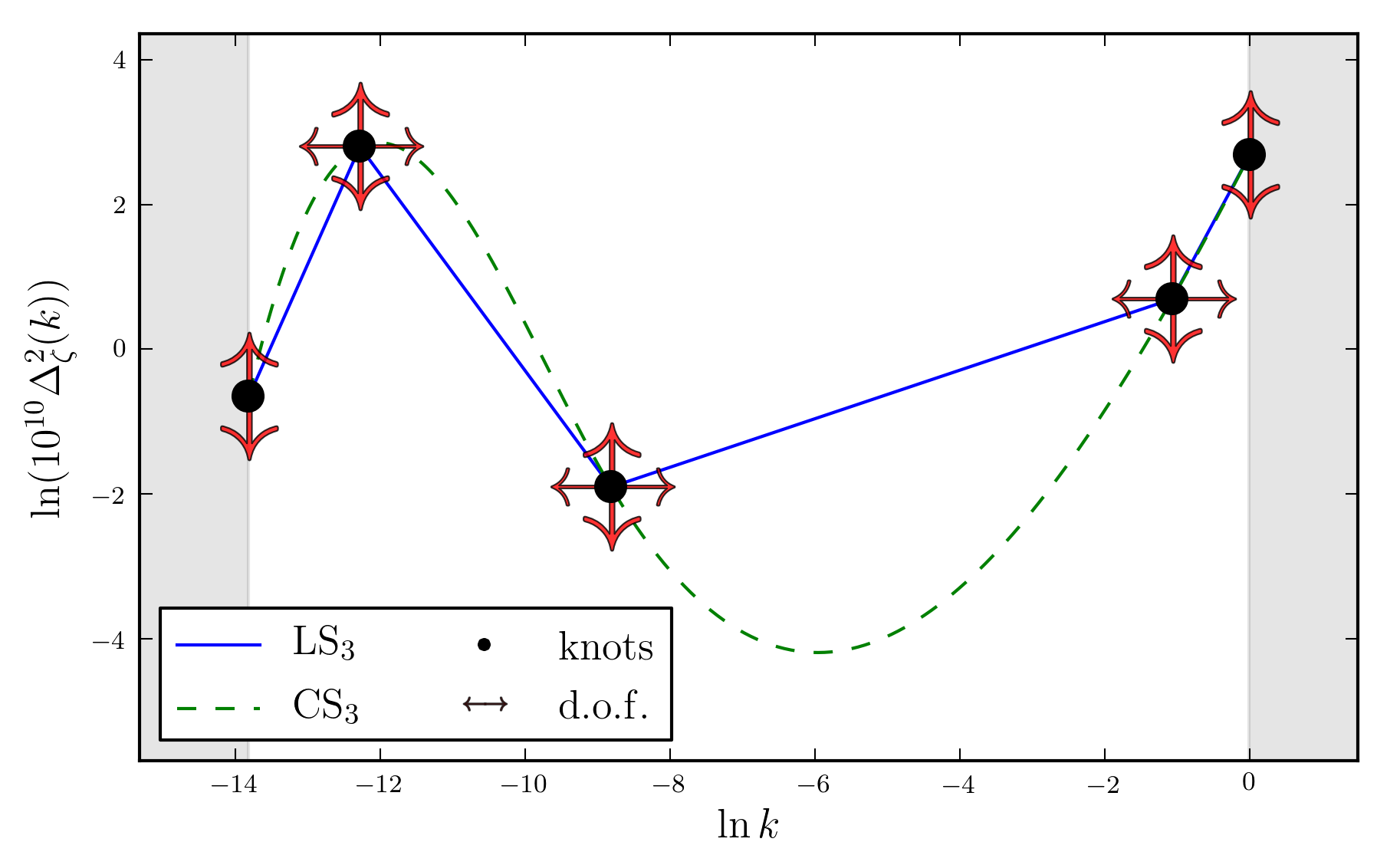}
\caption{We illustrate the linear-spline  $\mathrm{LS}_n$ and cubic-spline  $\mathrm{CS}_n$ models for the primordial power spectrum, where $n$ is the number of knots between the two endpoints. The white region denotes the range of $k$ for which the spectrum is defined, with  $k_\mathrm{min}=10^{-6}\,\mathrm{Mpc}^{-1}$ and $k_\mathrm{max}=1.0\,\mathrm{Mpc}^{-1}$.  There are $2n+2$ degrees of freedom for each choice of spline, since we vary the amplitude $\Delta^2_\zeta$ at the endpoints and allow the knots to move in both $\Delta^2_\zeta$ and $k$.}
\label{model_fig}
\end{figure}

Since no model-selection method can be completely non-parametric, we check our analysis by obtaining posterior probabilities for two different styles of non--power-law PPS.  We compare both a linear- and a cubic-spline interpolation model, which capture sharp and smooth features in the PPS, respectively.  The two models are illustrated in Fig.~\ref{model_fig} and explained in detail in Section~\ref{ssect:relaxingpowerlaw}.  We allow variation in the number of knots, their amplitudes, their positions in $k$-space, and the endpoint amplitudes. We see the maximum increase in evidence is $\Delta\ln Z=0.7$ for the one knot linear spline model with varying foregrounds and $\Delta\ln Z=2.2$ for the five knot linear spline, albeit with the foreground parameters in the Planck likelihood fixed to their best-fit values.

To test the effectiveness of our PPS parametrizations we attempt to recover nontrivial signals in simulated CMB temperature maps.  The method clearly finds even small added features in the PPS, while the evidence ratio strongly favors models with added interior knots when the simulated feature is large enough. The analysis here considers only \emph{Planck} data. In a separate paper we will consider the implications of the recent BICEP2 $B$-mode polarization detection \cite{Ade:2014xna,Ade:2014gua} for the scalar power spectrum using the knot-spline techniques developed here.

\section{Modeling the primordial power spectrum}\label{sec_methods}

\subsection{Bayesian evidence and model-selection}\label{ssect:bayes}

Cosmological data is intrinsically stochastic, so overfitting is a  key danger for empirical reconstructions of the PPS.  While many strategies have been suggested to prevent ``fitting to noise'',
the Bayesian approach naturally disfavors models that yield a small region with high likelihood in a much larger multidimensional parameter space. 

To compare two models $M_i$ and $M_j$ given data $D$, we evaluate the ratio of their posterior probabilities
\begin{equation}
  K_{ij} = \frac{P(M_i \, | \, D)}{P(M_j \, | \, D)} = \frac{P(M_i) Z_i(D)}{P(M_j) Z_j(D)} \, ,
  \label{eqn:posteriorratio}
\end{equation}
where $Z_i$ is the Bayesian evidence, or marginalized likelihood, given in terms of parameters $\theta$ and the data likelihood $\mathcal L_i (D \, | \, \theta) \equiv P (D \, | \, \theta, M_i)$ by
\begin{equation}
  Z_i(D) \equiv P(D \, | \, M_i) = \int  P(\theta \, | \, M_i) \mathcal L_i (D \, | \, \theta) d \theta.
  \label{eqn:evidence}
\end{equation}
The prior probabilities $P(M_i)$ and $P(\theta \, | \, M_i) \equiv P_i(\theta)$ incorporate our \emph{a priori} information about the model $M_i$ and the model's parameters $\theta$, respectively.  The evidence ratio or \emph{Bayes factor} $B_{ij}=Z_i/Z_j$, is the ratio of  the posterior model odds divided by the prior model odds and measures  the extent to which the data ``prefers'' $M_i$ over $M_j$.  If the model priors are  equal, $B_{ij}=K_{ij}$ and the Bayes' factor  directly measures the posterior model odds.   Generically, the integral in Eq.~\eqref{eqn:evidence} is numerically challenging, but it is rendered tractable by multimodal nested sampling. We implement this algorithm using the \textsc{MultiNest} package \cite{Feroz:2007fi,Feroz:2008eb,Feroz:2013uq}, which is widely employed for this purpose within cosmology.

Evidence ratios can be judged qualitatively using the Jeffreys' scale \cite{Jeffreys:1961} or a more conservative ``cosmology scale'' \cite{hobson2010bayesian}, collated in Table \ref{jeffreys_table}.  Negative evidences can be assessed by inverting the ratio before taking the logarithm in Eq. \eqref{eqn:posteriorratio}.   Bayesian model-selection is typically more conservative than frequentist hypothesis testing \cite{berger1991interpreting}.  For example, it is always possible to postulate a very complex PPS that will perfectly fit the data, thereby trivially maximizing $\mathcal L(D)$.
In fact, it is possible to obtain a significant improvement in $\chi^2$ or log-likelihood by letting the PPS have a lot of freedom \cite{2013JCAP...07..031H,2010JCAP...01..016N,2009JCAP...07..011N,2009PhRvD..79d3010N,2008PhRvD..78b3511S,2007PhRvD..75l3502S,2006MNRAS.367.1095T,2004PhRvD..70d3523S}, but this usually results in a  complex power spectrum with many curious wiggles and oscillations.
However, in these circumstances it is likely that the best fit  corresponds to a small subregion of the overall parameter space, and Bayesian evidence penalizes scenarios of this form.

While the Bayes factor quantifies the relative evidence between two models, it is not an ``all-purpose tool.'' Bayesian model selection is often characterized as an implementation of Occam's razor, as it penalizes models with many parameters.  In fact, a close inspection of Eq.~\eqref{eqn:evidence} shows that it  penalizes models that do not give sufficiently high likelihood over the integration volume.  If a model adds parameters $\theta'$ that are only weakly-constrained by the likelihood $\mathcal L (\theta') \sim 1$, the integrals over $\theta'$  in Eq.~\eqref{eqn:evidence}  factorise and cancel when evidence ratios are computed. Furthermore, if we make the common choice that the prior probabilities for the models are equal, Bayesian model selection is not testing against a ``null hypothesis'', but comparing two models; a  Bayes factor $|B_{ij}| \lesssim 1.0$ simply expresses the fact that the data used to construct the likelihood has no strong preference for either model.  The simpler model, \emph{i.e.}, the one with fewer parameters, is only preferred if we give it more \emph{a priori} weight.  In this paper we report both our quantitative calculations for $B_{ij}$ and a qualitative analysis of the posterior probabilities on the model parameters.

 While the Bayes factor $B_{ij}$ is one tool we can use to perform model-selection, our parameterization of the non--power-law model still plays a major role in the Bayesian evidence integral of Eq.~\eqref{eqn:evidence}.  An ideal non--power-law PPS parameterization needs to be able to construct any type of \emph{possible} deviation from a power-law PPS.  This is not practical, since this requires an infinite-dimensional parameter space.  There are too many possible forms for the PPS to explore and we inevitably must truncate the allowed degrees of freedom.  Instead, we have to settle for a model that can reconstruct a wide range of \emph{probable} PPS features.  This clearly depends on what features we expect to be present in the data, which introduces the possibility of confirmation bias into the analysis.

\begin{table}
\renewcommand{\arraystretch}{1.5}
\setlength{\arraycolsep}{5pt}
\centering
\begin{tabular}{c c | c c}
  $\ln K_{ij}$ & $K_{ij}$ & Jeffreys Scale & Cosmology Scale \\
\hline
0.0 to 1.0    & 1.0 to 2.7    & \multicolumn{2}{c}{Not worth more than a bare mention}  \\
1.0 to 2.5 & 2.7 to 12.2 & Substantial & Weak \\
2.5 to 5.0 & 12.2 to 148.4 & Strong & Significant \\
$> 5$      & $>148.4$ & Decisive & Strong
\end{tabular}
\caption{Rough guideline for Bayesian evidence interpretation with the Jeffreys scale \cite{Jeffreys:1961} and the re-scaled ``cosmology scale'' from Ref. \cite{hobson2010bayesian}. Assuming two models $\mathcal M_i$ and $\mathcal M_j$ have the same \emph{prior probability}, the Bayes factor $B_{ij}$ is equivalent to the ratio of posterior probabilities $K_{ij}$ and can be interpreted directly as the \emph{posterior betting odds} for $\mathcal M_i$ over $\mathcal M_j$.  The more conservative scale emphasizes that in cosmology there is often substantial uncertainty in the choice and form of model priors and the resulting evidences should therefore be interpreted more carefully.}
\label{jeffreys_table}
\end{table}

\subsection{The power-law power spectrum}

As reviewed in Ref.~\cite{Baumann:2009tu}, the primordial scalar perturbations can be expressed in terms of the gauge invariant curvature perturbations $\zeta(\mathbf{x})$ in real space. In Fourier space they become
\begin{equation}
\zeta(\mathbf{k})=\int\,d^3x\,e^{-i\mathbf{k}\cdot\mathbf{x}}\zeta(\mathbf{x})\,.
\end{equation}
If the perturbations are statistically homogeneous, the two-point correlation function is
\begin{equation}
\vev{\zeta(\mathbf{k})\zeta^*(\mathbf{k^\prime})}=(2\pi)^3\delta^3(\mathbf{k}-\mathbf{k^\prime})
\left(\frac{2\pi^2}{k^3}\Delta^2_\zeta(k) \right)
\label{eqn:twopoint}
\end{equation}
where $\delta^3$ denotes the delta-function in three dimensions and $\Delta^2_\zeta$ is the dimensionless {\it primordial power spectrum (PPS)}.

If the perturbations are Gaussian, they are fully described by Eq. \eqref{eqn:twopoint}. Given that no primordial non-Gaussianity has been detected  \cite{Collaboration:2013ww,Collaboration:2013wc} we  restrict our analysis to the Gaussian case.  In the standard $\Lambda$CDM model the PPS is described by a simple power law  with  two free parameters
\begin{equation}\label{pps_standard}
\Delta^2_\zeta(k)=A_s\left(\frac{k}{k_*}\right)^{n_s-1} \, ,
\end{equation}
where $A_s$ is the {\it amplitude} of the scalar perturbations; $n_s$ is  the {\it scalar spectral index}; and $k_*$ is the {\it pivot scale}. The latest CMB temperature data from \emph{Planck} \cite{Collaboration:2013ww}, CMB polarization data from WMAP \cite{Bennett:2012fp}, small scale temperature data from ACT \cite{Sievers:2013wk} and SPT \cite{Story:2012wx}, as well as the large scale structure data are all in excellent agreement with this assumption. These data sets in combination \cite{Collaboration:2013ww} put the following bounds on the PPS parameters: $\ln(10^{10}A_s)=3.091\pm0.025$, $n_s=0.9608\pm0.0054$ at $k_* = 0.05 \, \impc$.

\subsection{Relaxing the power-law}
\label{ssect:relaxingpowerlaw}

The  goal of this paper is to explore the consequences of relaxing the assumption expressed by Eq.~\eqref{pps_standard}. We wish to let the PPS have a model independent form and  reconstruct it from  experimental data, and  implement a version of the ''knot-spline'' reconstruction method developed in \cite{2012JCAP...06..006V}. This approach is described qualitatively in in Fig.~\ref{model_fig}. Specifically, we assume that $P(\log{k})$ is either a linear or cubic spline with $n$ knots, defined between fixed endpoints.  In the absence of knots, this model is functionally equivalent to a power-law, given that we are interpolating in $\log{k}$. We systematically add knots to allow for more complex features.\footnote{The simplest case with only one parameter, \emph{i.e.}, a constant PPS is the familiar Harrison-Zel'dovich spectrum, and is disfavored by \emph{Planck} at  $>5\sigma$ \cite{Collaboration:2013ww}.}   

The spectrum is generated via the following algorithm:
\begin{enumerate}
  \item We fix $k_\mathrm{min} =10^{-6}\,\mathrm{Mpc}^{-1}$ and $k_\mathrm{max}=1.0\,\mathrm{Mpc}^{-1}$, allowing the PPS to vary only in amplitude $\Delta_\zeta^2(k)$ at the endpoints.

  \item Add $n$ knots, chosen with a uniform prior on $\ln k$, in the range $\ln k_\mathrm{min} < \ln k_i < \ln k_\mathrm{max}$ and a uniform prior on $-2 <  A_i < 4$ for $A_i \equiv \ln(10^{10}\Delta^2_\zeta(k_i))$, with $i=1,2,...,n$.  We then order the set of knots so that $k_{i-1} \le k_i$.

  \item Interpolate between the  endpoints and the $n$ ordered knots by  a linear spline ($\mathrm{LS}_n$) or cubic spline ($\mathrm{CS}_n$).  Perform the interpolation in logarithmic space for both $k$ and $\Delta^2_\zeta$.

\end{enumerate} 

The endpoints adopted here define an overall range of $k$ that maps to angular scales from  $l=2$ to $l\approx14000$, with the rough relationship $l\sim kL_0$, where $L_0\approx14.4\,\mathrm{Gpc}$ is the distance to the last scattering surface.  The high-end of this range is well past the $l$-values accessible by \emph{Planck} or any other current or near-future experiment.  However, this range coincides with the convention typically used in the Boltzmann-solvers CLASS \cite{Lesgourgues:2011ud,Blas:2011vf} and CAMB \cite{Lewis:1999bs}, even when $C_l$'s less than $l\sim2000$ are not evaluated. The logarithmic priors on the knots $k_i$ and their amplitudes $A_i$ indicate that these is no preference for the amplitude and location of any feature(s) and  knots with arbitrary locations in $k$-space can capture both local features and gradual modulations in the PPS.

As Fig.~\ref{model_fig} shows, different choices for the interpolation scheme between knots can result in wildly different power spectra.  The linear-spline interpolation ($\mathrm{LS}_n$) constructs a continuous but not differentiable PPS by connecting the endpoints and knots by simple line segments. On the other hand, the cubic spline interpolation ($\mathrm{CS}_n$) constructs a PPS with continuous first and second derivatives, connecting knots and endpoints with segments of cubic polynomials.  This lets us use the knot-interpolation approach to model and reconstruct many possible features, although a $P(k)$ with many turning points (such as a rapid modulation) would require a prohibitive number of knots.  The linear-spline technique is best suited to sharp transitions in the PPS, while the cubic-spline models smooth deviations from the simple power law case.\footnote{The PPS is numerically represented as a cubic spline of a large number of sample points for either case. This means that the sharp corners of the linear spline interpolation will be smoothed out. This helps avoid numerical problems that could arise because of the discontinuity of the first derivative of the PPS for the linear spline case.} Since the positions of the knots are not fixed, they can recover step-like features anywhere in $k$-space, especially with $\mathrm{LS}_n$, as well as a  cut-off  PPS  (e.g., Ref. \cite{Bridges:2008ta}).

We have stipulated $k_{i-1} \le k_i$, so the volume of parameter space corresponding to the knot locations is 
\begin{equation}
  V_n= \frac{1}{ n!} \left(\Delta \ln k \right)^n \, .
  \label{eqn:XXX}
\end{equation}
If we had not rank-ordered the knot positions we would have $n!$ combinations of $k_i$ and the corresponding parameter region would be an $n$-cube with volume $n! V_n$. However, in this case the marginalized likelihood integral  splits into a sum of $n!$ identical terms after appropriate relabelings of the $k_i$ and the two factors of $n!$  cancel. Consequently, the evidence values do not depend on whether the $k_i$ are assumed to be ordered, but the posterior distributions of the individual $k_i$ will  depend strongly on this choice.

With no knots ($n=0$) the PPS is described by only two parameters: the amplitudes of the PPS at the endpoints $k_\mathrm{min}$ and $k_\mathrm{max}$.  In this limit, both choices $\mathrm{LS}_0$ and $\mathrm{CS}_0$ are equivalent to the power-law PPS of Eq.~\eqref{pps_standard}. This makes it very easy to compare the Bayesian evidence for the primordial power spectra with  features to the standard power law case.  Each additional knot adds two extra degrees of freedom to the model, its location $k_i$ and its amplitude $A_i$, yielding $2n+2$ total parameters in each PPS parameterization.

\section{Data, likelihoods and basic checks}\label{data_like_sec}

The remaining  cosmological parameters $\Omega_bh^2$, $\Omega_ch^2$, $h$, and $\tau$ have uniform priors in the ranges $[0.020, 0.025]$, $[0.1, 0.2]$, $[0.55, 0.85]$, and $[0.02, 0.20]$, respectively. Here, $\Omega_b$ denotes the dimensionless baryon density, $\Omega_c$ is the dimensionless cold dark matter density, $h$ defines the Hubble parameter $H$ via $H=100\,\mathrm{km}/\mathrm{s}/\mathrm{Mpc}$, and $\tau$ is the reionization optical depth.

We use the publicly available code \textsc{Cosmo++} \cite{Aslanyan:2013ts} for our analysis. The CMB power spectra are calculated using the \textsc{CLASS} package \cite{Lesgourgues:2011ud,Blas:2011vf} and we use the multimodal nested sampling, implemented in the publicly available code \textsc{MultiNest} \cite{Feroz:2007fi,Feroz:2008eb,Feroz:2013uq}, to compute posteriors and Bayesian evidence. The publicly available \emph{Planck} likelihood code \cite{Collaboration:2013vc} is used for the PPS reconstruction in Section \ref{sec_pps}.  We use the high-$l$ \textsc{CamSpec} likelihood, the low-$l$ \textsc{Commander} likelihood, and the lensing likelihood, thus incorportating all available  \emph{Planck} data, but do not include information from any other sources.  The hemispherical analysis in Section \ref{sec_hemispheres}  uses the SMICA map \cite{Collaboration:2013vx} and the  \textsc{Cosmo++} likelihood code, which is also used in Section \ref{sec_simulations} for analyzing simulated maps. The analysis presented here was carried out using  \textsc{Cosmo++} package \cite{Aslanyan:2013ts}, and the knot-spline PPS is available in the current version of this package.  

We  test our tool-chain by estimating parameters for  the power-law PPS, Eq.~\eqref{pps_standard}, using the \emph{Planck} likelihood code with lensing. We find excellent agreement with the corresponding \emph{Planck}  constraints:\footnote{In particular, compare these results to those of Table 9 in Ref. \cite{Collaboration:2013ww}.} $\Omega_bh^2=0.02219\pm0.00032$, $\Omega_ch^2=0.1185\pm0.0030$, $h=0.684\pm0.015$, $\tau=0.089\pm0.030$, $n_s=0.9625\pm0.0090$, $A_s=3.085\pm0.054$. With no knots  the $\mathrm{LS}_0$ and the $\mathrm{CS}_0$ are both equivalent to the standard case and, as expected, the parameter constraints for these models are almost indistinguishable from the constraints given above.

The \emph{Planck} likelihood code includes $14$ nuisance parameters, which mostly relate to unmodelled foregrounds  \cite{Collaboration:2013vc}. For the $\mathrm{LS}_n$ scenarios we have done the analysis with the foreground parameters both varying and fixed to the central values with a power law PPS. We have verified that the shapes of the posteriors do not change significantly when the foreground parameters are fixed. The computational cost of our procedure is substantially higher for the cubic spline than the linear spline and for convenience the cubic spline analysis was performed with the nuisance parameters fixed. 

\section{Should we trust our reconstruction?}\label{sec_simulations}

\begin{figure}
\centering
\includegraphics[width=12cm]{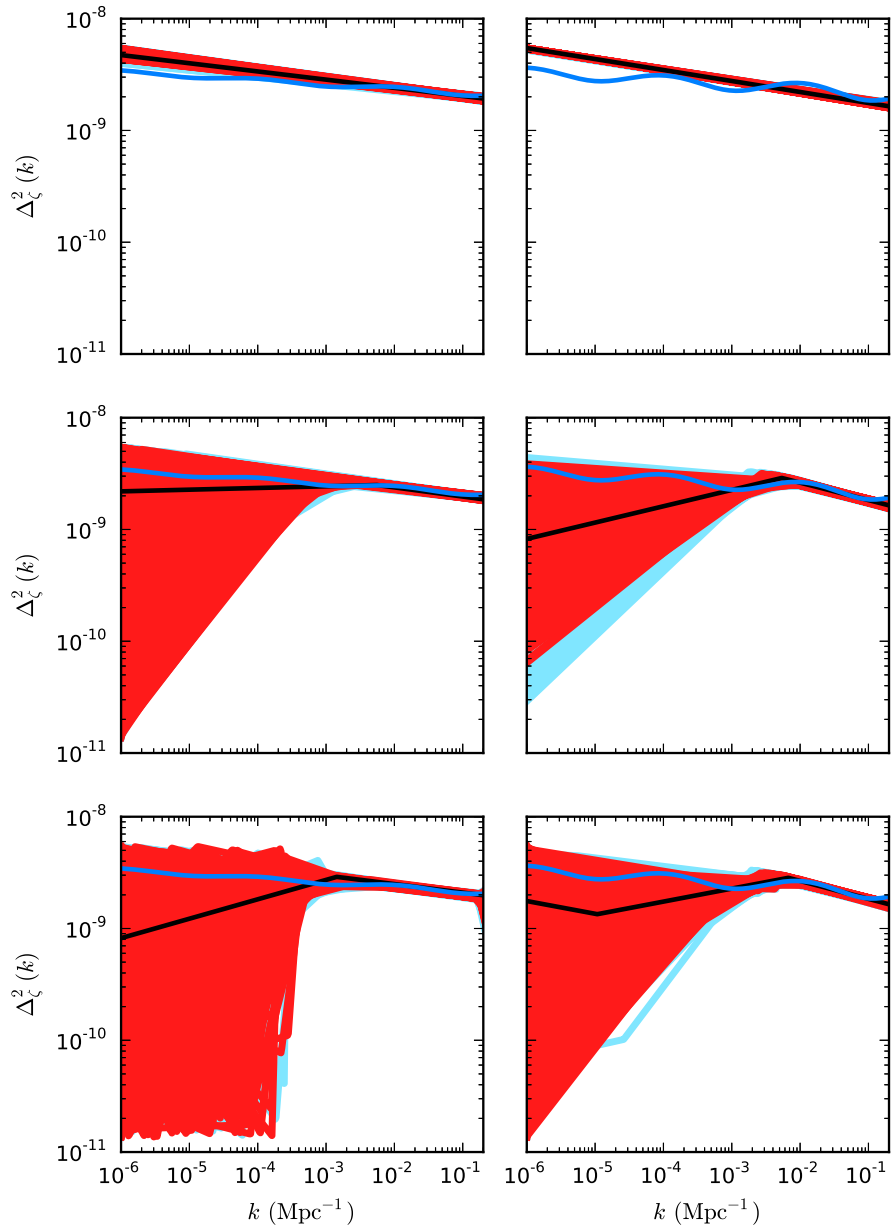}
\caption{\label{ps_sim_lin_fig} The reconstructed primordial power spectrum (PPS) with simulated data. The columns correspond to (\emph{left}) $A^\prime_s=1\times10^{-10}$ and (\emph{right}) $A^\prime_s=3\times10^{-10}$. The number of knots increases from $0$ to $2$ from top to bottom. The black solid lines show the best-fit PPS, the red lines are the PPS in the $68\%$ CI, and the light blue lines are the PPS in the $95\%$ CI. The blue solid line shows the modified PPS used for the simulation. Linear spline interpolation is used for all of the plots.
}
\end{figure}

Before we apply our approach to \emph{Planck} data, we investigate its ability to recover a non-power-law PPS from a temperature map with an injected non--power-law PPS signal in the form of a sinusoidal modulation to the standard power-law PPS
\begin{equation}\label{modified_pps}
\Delta^2_{\zeta,\mathrm{sim}}(k)=A_s\left(\frac{k}{k_*}\right)^{n_s-1}+A^\prime_s\cos\left(\frac{2\pi \ln k}{\ln\lambda^\prime}\right)\,.
\end{equation}
For simplicity we use the standard $\Lambda$CDM parameters at \emph{Planck} best-fit values: $\Omega_bh^2=0.022032$, $\Omega_ch^2=0.12038$, $h=0.6704$, $\tau=0.0925$, $A_s=2.2154\times10^{-9}$, $n_s=0.9619$, $k_*=0.05\,\mathrm{Mpc}^{-1}$. For the modification, we use the wavelength $\lambda^\prime=0.01\,\mathrm{Mpc}$ and two different values for the amplitude: $A^\prime_s=1\times10^{-10}$ and $A^\prime_s=3\times10^{-10}$.  

\begin{table}
\renewcommand{\arraystretch}{1.5}
\setlength{\arraycolsep}{5pt}
\begin{eqnarray*}
\begin{array}{c|cc}
n & A^\prime_s=1\times10^{-10} & A^\prime_s=3\times10^{-10} \\
\hline
0 & 0.0 & 0.0 \\
1 & 0.3 & 51.0 \\
2 & 0.8 & 51.7
\end{array}
\end{eqnarray*}
\caption{Change in Bayesian evidence $\Delta\ln Z$ with respect to $\Lambda$CDM as a function of the number of knots for two simulated maps.  The maps have an artificial sinusoidal primordial power spectrum of amplitude $A_s'$. The errors on $\ln Z$ are smaller than $\pm0.3$ in all cases.}
\label{sim_lin_table}
\end{table}

Our simulated maps have similar characteristics  to the SMICA map from Planck \cite{Collaboration:2013vx}. We set the \textsc{HEALPix} parameter $N_\mathrm{side}=2048$ and add white noise to the simulated maps which is similar to the instrumental noise in SMICA. We combine the low-$l$ pixel space likelihood code in the \textsc{COSMO++} package \cite{Aslanyan:2013ts} with the high-$l$ likelihood code in $C_l$ space. The pixel space code is used for $l=2\rightarrow30$ and the $C_l$-space code is used for $l=31\rightarrow1750.$\footnote{The simulated maps become noise dominated after $l\approx1750$,  like the SMICA map.} For the pixel space analysis we reduce the resolution of the simulated maps to \textsc{HEALPix} $N_\mathrm{side}=16$ by first transforming the maps to harmonic space with the {\tt anafast} routine from \textsc{HEALPix}, then smooth with a Gaussian kernel with a FWHM of $10^\circ$ using the {\tt alteralm} routine, and finally transform back to pixel space with a lower resolution using {\tt synfast}. For the high-$l$ likelihood calculation we first derive the $C_l$ values from the data using the implementation of the MASTER algorithm \cite{Hivon:2001eh} in \textsc{COSMO++} \cite{Aslanyan:2013ts}. We combine the SMICA, NILC, and SEVEM masks by Planck to get an approximation of the U73 mask, which is not publicly available. For the pixel space analysis we reduce the resolution of the mask by first smoothing it with a Gaussian kernel with a FWHM of $10^\circ$, then downgrading it with the {\tt ud\_grade} routine from \textsc{HEALPix}, then masking out all the pixels with a final value lower than $0.8$. 

We apodize the combined mask with a $30^\prime$ cosine function for the high-$l$ $C_l$ calculation. We use this mask for the simulated maps to make the analysis similar to the real SMICA map analysis. The priors on all of the parameters are the same as in Section~\ref{data_like_sec}, except for the optical depth of reionization $\tau$. Since no polarization or lensing data is included in this analysis, we use an informative prior of $\tau=0.0851\pm0.014$.  This follows the analysis of foreground cleaned temperature maps for \emph{Planck} \cite{Collaboration:2013vx}.

The posterior probability distributions for the knot positions, amplitudes, and the best-fit PPS, are shown in Fig.~\ref{ps_sim_lin_fig}.  With no knots (equivalent to the power law PPS) the best fit spectrum is a good match to the simulated spectrum at small scales, \emph{i.e.}, $k \, \gsim \, 10^{-3} \, \impc$. With $A_s'=1\times10^{-10}$ we recover a power-law with $n_s=0.924$, while for $A_s'=3\times10^{-10}$ we get $n_s=0.900$.  The added modulation increases the tilt on small scales,  making both spectra redder than \emph{Planck}'s.  With a single  knot we detect the local maximum in Eq.~\eqref{modified_pps} at $k\approx10^{-2}\, \impc$ or $l\sim150$.  Adding more knots makes less qualitative difference to the posterior distributions as there is only one local maximum on small scales.  However,  models with knots give posterior probabilities that indicate a suppression of the PPS for $k \lesssim 10^{-3} \, \impc$ for both simulations, although the assumed PPS in Eq.~\eqref{modified_pps} has no such feature.  The model reconstructs the slight depression at $10^{-3} \, \impc \lesssim k \lesssim 10^{-2} \, \impc$, but is not able to capture the small-$k$ behavior, due to cosmic variance.  Consequently, a preference for PPS suppression at $k \lesssim 10^{-3} \, \impc$ (or $l \lesssim 15$) can be degenerate with a strong, local feature at scales that are better constrained by the likelihood.

 Table~\ref{sim_lin_table}  shows  Bayesian evidence ratios for the simulated maps.  For a very mild modulation ($A^\prime_s=1\times10^{-10}$)  Bayesian evidence is not able to distinguish between  models with $0$, $1$, or $2$ knots. However, for a stronger modulation ($A^\prime_s=3\times10^{-10}$) adding a single knot yields a large increase in evidence, $\Delta\ln Z = 51$,    conclusively favoring the one knot reconstruction. However models with two or more knots do not give significant extra improvement, due to cosmic variance on large scales.

 These simulated maps show that our method can detect features in the power spectrum, even if the features are relatively small.  Bayesian evidence clearly distinguishes between models with and without features for the case of the large modulation.   However, we have deliberately chosen an uninformative prior, which allows a weakly constrained search but dilutes the ability of the evidence calculation to confirm the presence of smaller modulations.\footnote{See Ref.~\cite{Easther:2011yq} for a discussion of this problem in the context of inflationary model selection.} Finally, we must be careful when interpreting posterior probabilities for the PPS that show a decrease in power for $k<10^{-3} \, \impc$ as this suppression can be mimicked by a local feature.

\section{Reconstructing with \emph{Planck}}\label{sec_pps}

\subsection{Primordial power spectrum}

\begin{figure}
\centering
\includegraphics[width=13.5cm]{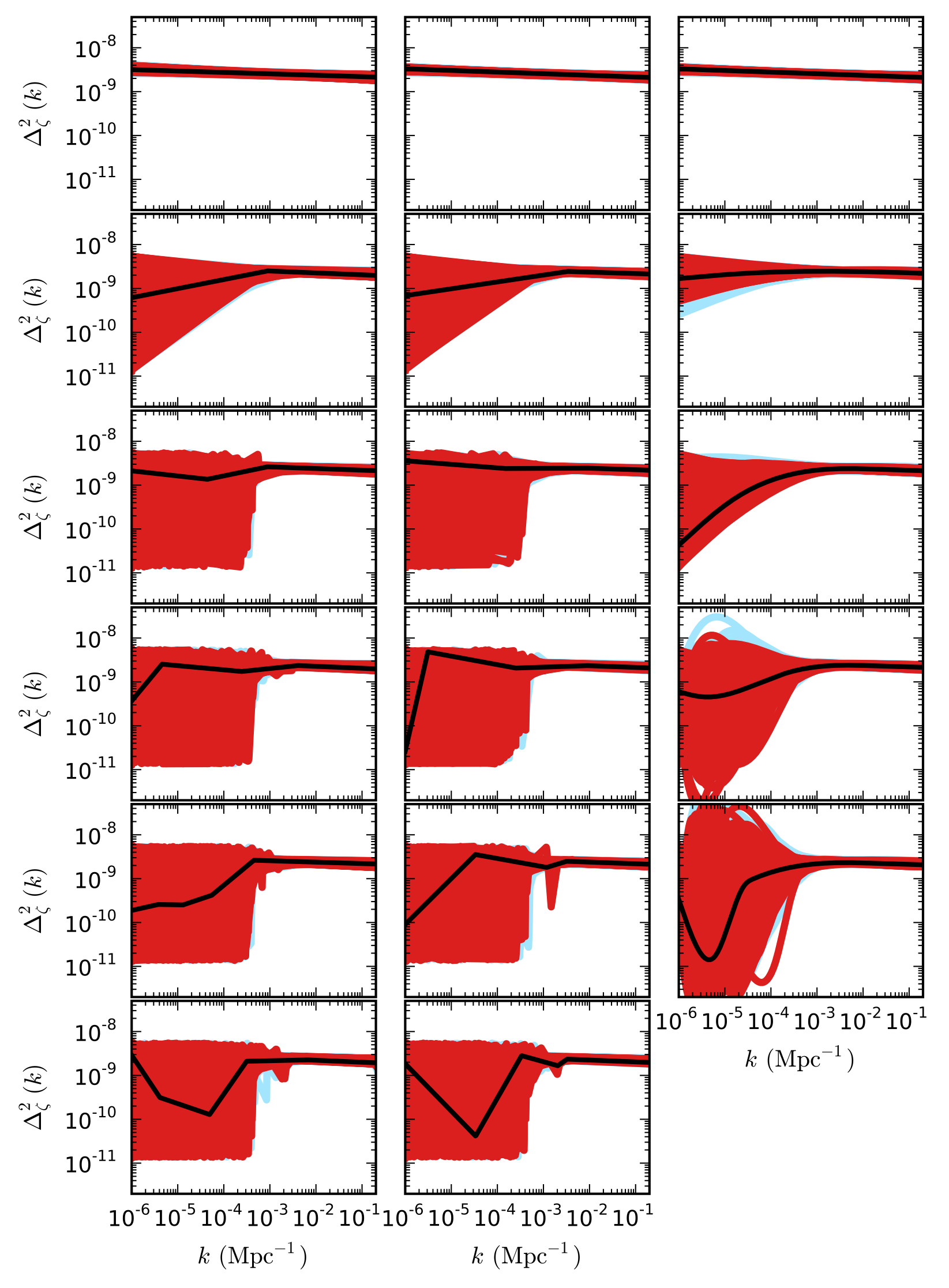}
\caption{\label{pps_planck_fig} The reconstructed primordial power spectrum (PPS) from the \emph{Planck} data. The columns correspond to (\emph{left}) linear-spline interpolation with $n$ knots ($\mathrm{LS}_n$), while varying the foreground cosmology parameters; (\emph{middle}) $\mathrm{LS}_n$ with fixed foreground parameters; and (\emph{right}) cubic-spline interpolation with fixed foreground parameters. The number of knots increases from $0$ to $5$ from top to bottom. The black solid lines show the best-fit PPS, the red lines are the PPS in the $68\%$ CI, and the light blue lines are the PPS in the $95\%$ CI.
}
\end{figure}

\begin{figure}
\centering
\includegraphics[width=13.75cm]{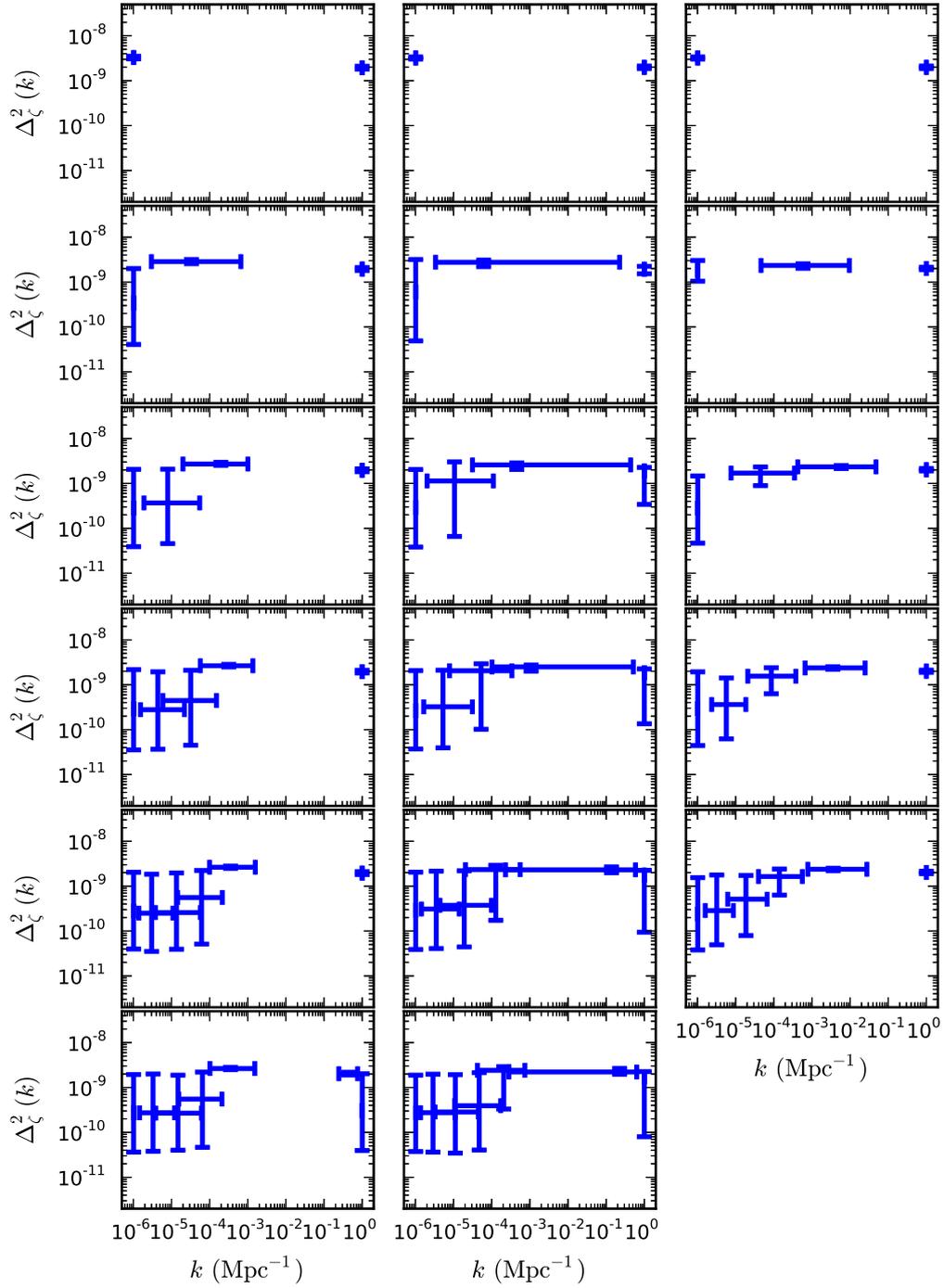}
\caption{Constraints on the location of knots for the non--power-law primordial power spectrum at the $68\%$ CI.  There are $n$  knots, whose location varies in $k_i$ and $\Delta^2_\zeta$, and two endpoints that vary only in $\Delta^2_\zeta$.   The columns are: (\emph{left}) a linear-spline ($\mathrm{LS}_n$) and varying foreground parameters; (\emph{middle}) $\mathrm{LS}_n$ with fixed foreground parameters; and (\emph{right}) cubic-spline ($\mathrm{CS}_n$) with fixed foreground parameters. The number of  knots increases from $0$ to $5$ from top to bottom.}
\label{knot_bounds_fig}
\end{figure}

\begin{figure}
\centering
\includegraphics[width=10cm]{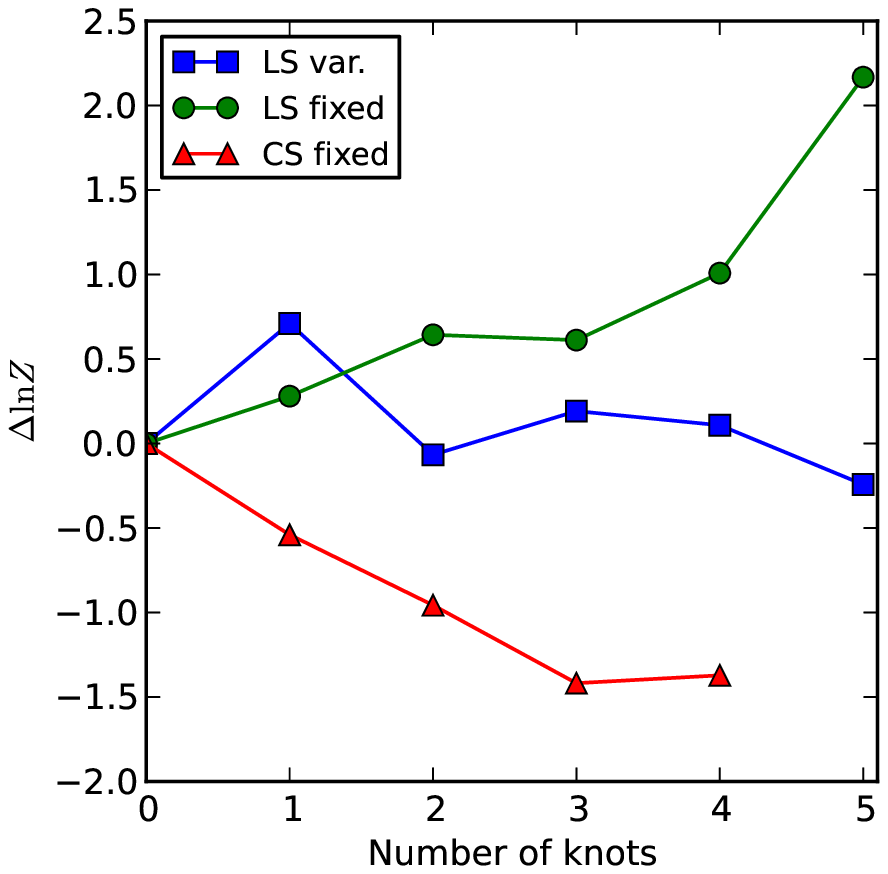}
\caption{\label{pps_planck_evidence_fig} Change in Bayesian evidence $\Delta \ln Z$ with respect to $\Lambda$CDM as a function of the number of knots in the primordial power spectrum model. The blue squares correspond to a linear-spline ($\mathrm{LS}_n$) without fixing the foreground parameters; the green circles correspond to $\mathrm{LS}_n$ with fixed foreground parameters; and the red triangles correspond to a cubic-spline ($\mathrm{CS}_n$) with fixed foreground parameters.
}
\end{figure}

Figure~\ref{pps_planck_fig} gives the posterior probability distributions for the ``knot-spline'' primordial power spectrum (PPS) reconstruction.  We use the linear-spline ($\mathrm{LS}_n$) model with up to $n=5$  knots and the cubic-spline ($\mathrm{CS}_n$) model with up to $n=4$  knots\footnote{The cubic-spline reconstruction is computationally more expensive. For this reason we restrict our analysis with this model up to $4$  knots.} and show the best-fit PPS in black.  The $68\%$ and $95\%$ CI are shown in red and light blue, respectively.  The constraints on the positions of the knots are given in Fig.~\ref{knot_bounds_fig} and the numerical values are given explicitly in Table~\ref{knot_bounds_table} of Appendix~\ref{appendix}.

None of the posteriors show any features beyond $k \gtrsim 2\times10^{-3}$ (roughly $l \gtrsim 30$), which is the region best-constrained by the \emph{Planck} likelihoods.  The position of maximum likelihood for the last knot is relatively stable at $k \sim 5\e{-3} \, \impc$, implying there is little evidence for global features.  Furthermore, we do not see any pair of knots (or knot/endpoint) that consistently retains its position as more knots are added, indicating that there are no localized features in the PPS for scales that are well-constrained by data.

However, most of the reconstructed power spectra give preference to suppressed power below $k \lesssim 10^{-3} \, \impc$, in agreement with previous results \cite{2012JCAP...06..006V,Wang:2013vf}. The cosmic variance, however, is largest on these large scales.  Furthermore, as described in Section~\ref{sec_simulations}, even if there were significant evidence in favor of a non--power-law model, this low-$k$ suppression should be treated with care, since it could also indicate a local feature near $k\sim10^{-3} \, \impc$.

To evaluate the integrated likelihood for the non--power-law models, we report the Bayesian evidence in Fig.~\ref{pps_planck_evidence_fig}.  We use the Bayesian evidence for the PPS models with no knots as a reference, since this case corresponds to the standard power-law PPS of Eq.~\eqref{pps_standard} and the $\Lambda$CDM model. There is a slightly increased evidence for the linear-spline model with one knot ($\mathrm{LS}_1$), both with and without varying the foreground parameters, with a maximum of $\Delta\ln Z = 0.7$. For more knots, the Bayesian evidence is smaller than the power-law PPS for the cubic-spline model and the linear-spline model with varying foreground parameters.  The $\mathrm{CS}_n$ reconstruction gives smaller Bayes factors compared to the standard case for any $n$.
When a linear-spline interpolation is used with fixed foreground parameters, we see a slightly increased evidence for a higher numbers of knots.  However, in this case we are ignoring the impact of marginalising over the foreground parameters, which introduces a nontrivial uncertainty into Bayesian evidence calculation.  When the foreground parameters are varied, this evidence does not increase with $n$.  Therefore, we conclude that we see no significant preference for a higher number of knots.

A frequentist search for broad features in the power spectrum using  Planck temperature data was   performed in Ref.~\cite{2013JCAP...12..035H}. The authors  used binned power spectra with a fixed value of $n_s$ with up to four bins, and a varying $n_s$ with up to two bins. Their two-bin Model-C reconstruction is equivalent to our linear spline reconstruction with $1$ knot. Our best-fit $\mathrm{LS}_1$ power spectrum agrees very well with the best-fit case of \cite{2013JCAP...12..035H} and the overall conclusions of \cite{2013JCAP...12..035H} are consistent with with our Bayesian results.

\begin{figure}
\centering
\includegraphics[width=15cm]{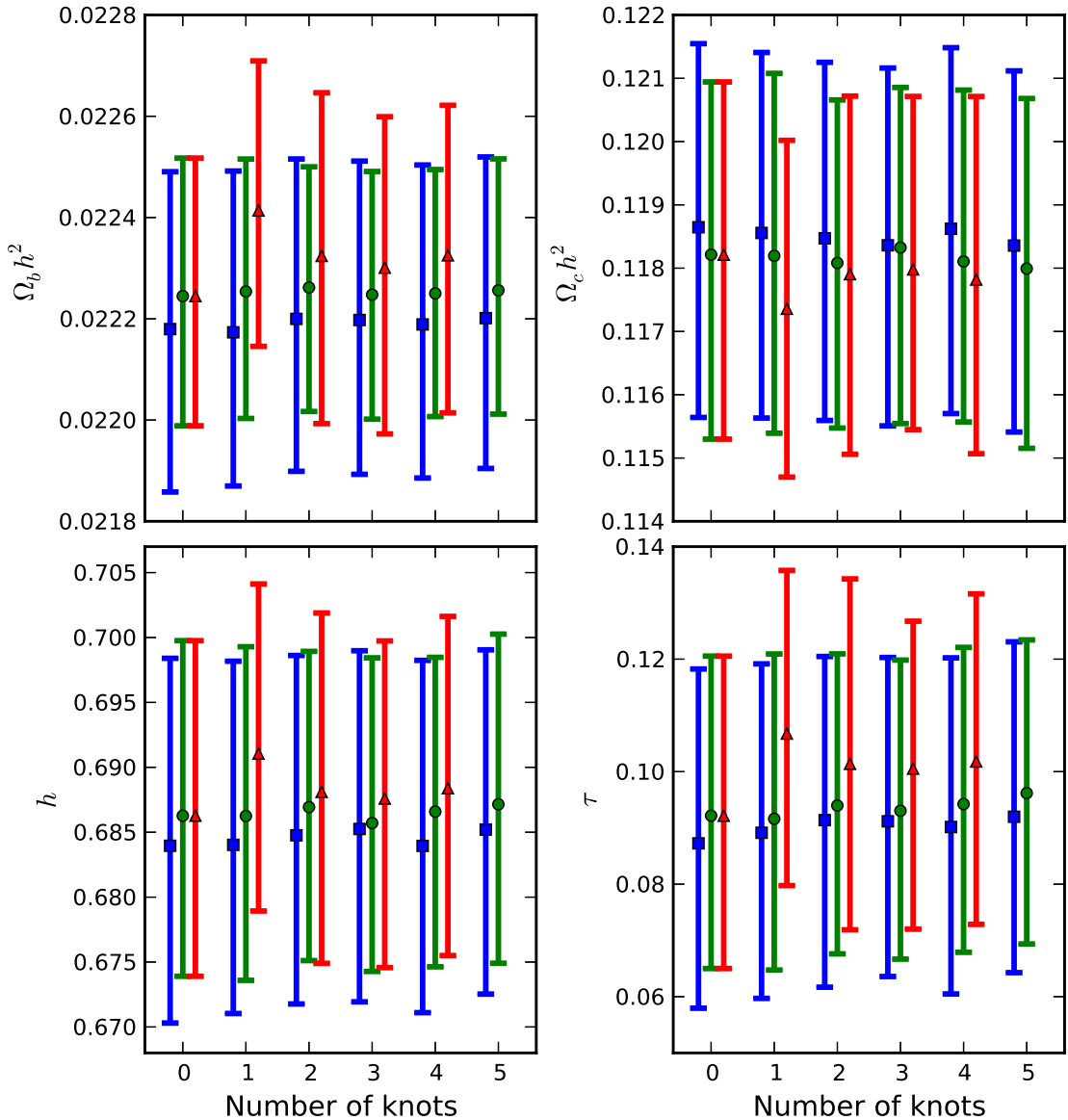}
\caption{\label{cosmo_params_fig} Cosmological parameter constraints as a function of the number of  knots $n$ at the $68\%$ CI. The blue squares correspond to a power spectrum linear-spline interpolation and varying foreground parameters, the green circles correspond to a linear-spline interpolation with fixed foreground parameters, and the red triangles represent the results with a cubic-spline interpolation and fixed foreground parameters.
}
\end{figure}

Figure~\ref{cosmo_params_fig} presents the posterior probabilities for the standard parameters $\Omega_bh^2$, $\Omega_ch^2$, $h$, and $\tau$ for both the linear-spline ($\mathrm{LS}_n$) and cubic-spline ($\mathrm{CS}_n$) models. The constraints on the parameters do not change significantly when more freedom is given to the PPS. The constraints with fixed foreground parameters are in good agreement with those found when they are free to vary; the error bars shrink slightly with fixed foreground parameters, as expected.  We do not see any significant changes in the constraints of the cosmological parameters with up to $5$ knots in the PPS, implying that the parameter constraints by \emph{Planck} \cite{Collaboration:2013ww} are robust and do not depend sensitively on the strong power-law assumption on the form of the PPS.

\subsection{Hemispherical power asymmetry}\label{sec_hemispheres}

\begin{figure}
\centering
\includegraphics[width=7.5cm]{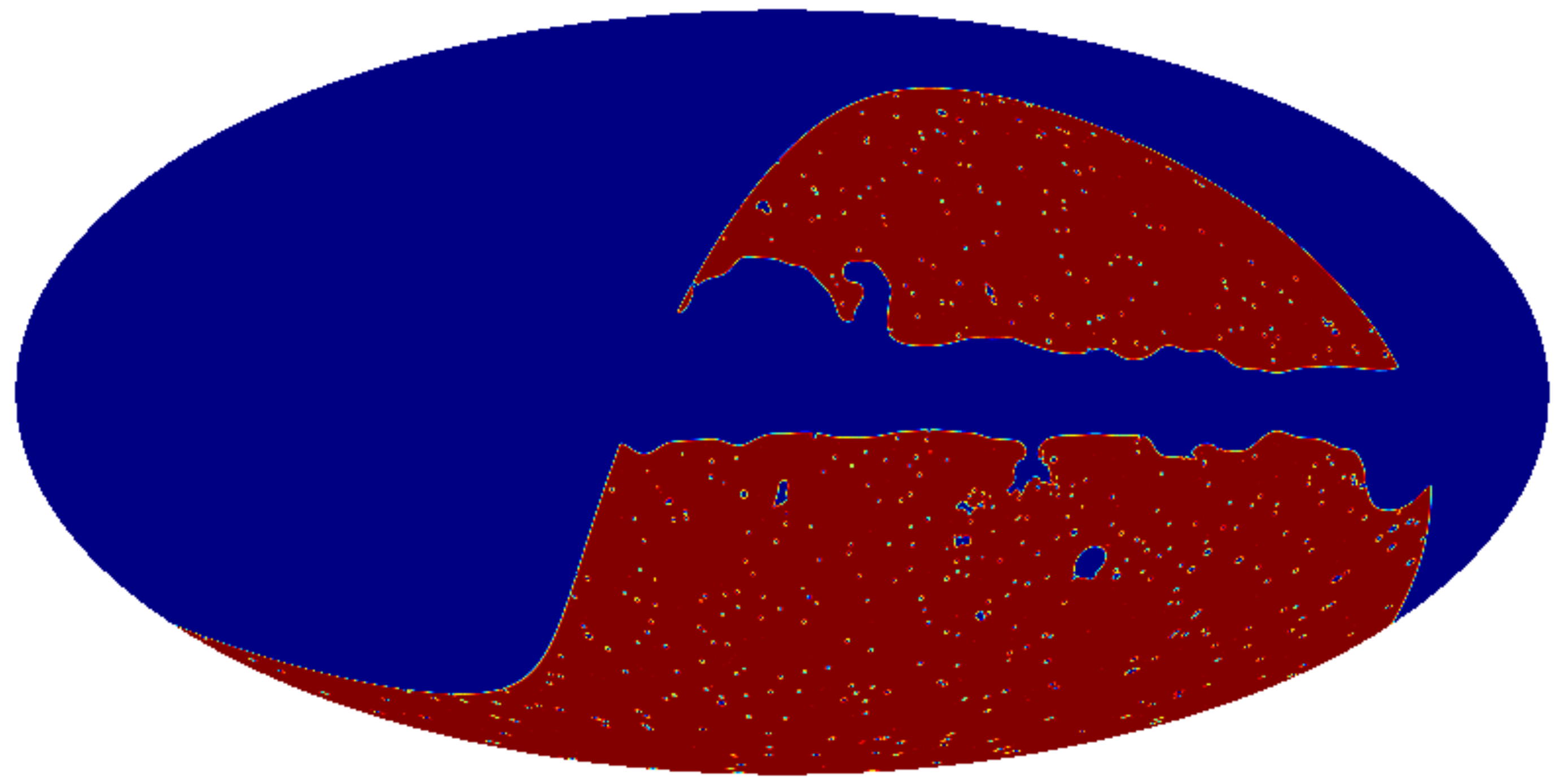}
\includegraphics[width=7.5cm]{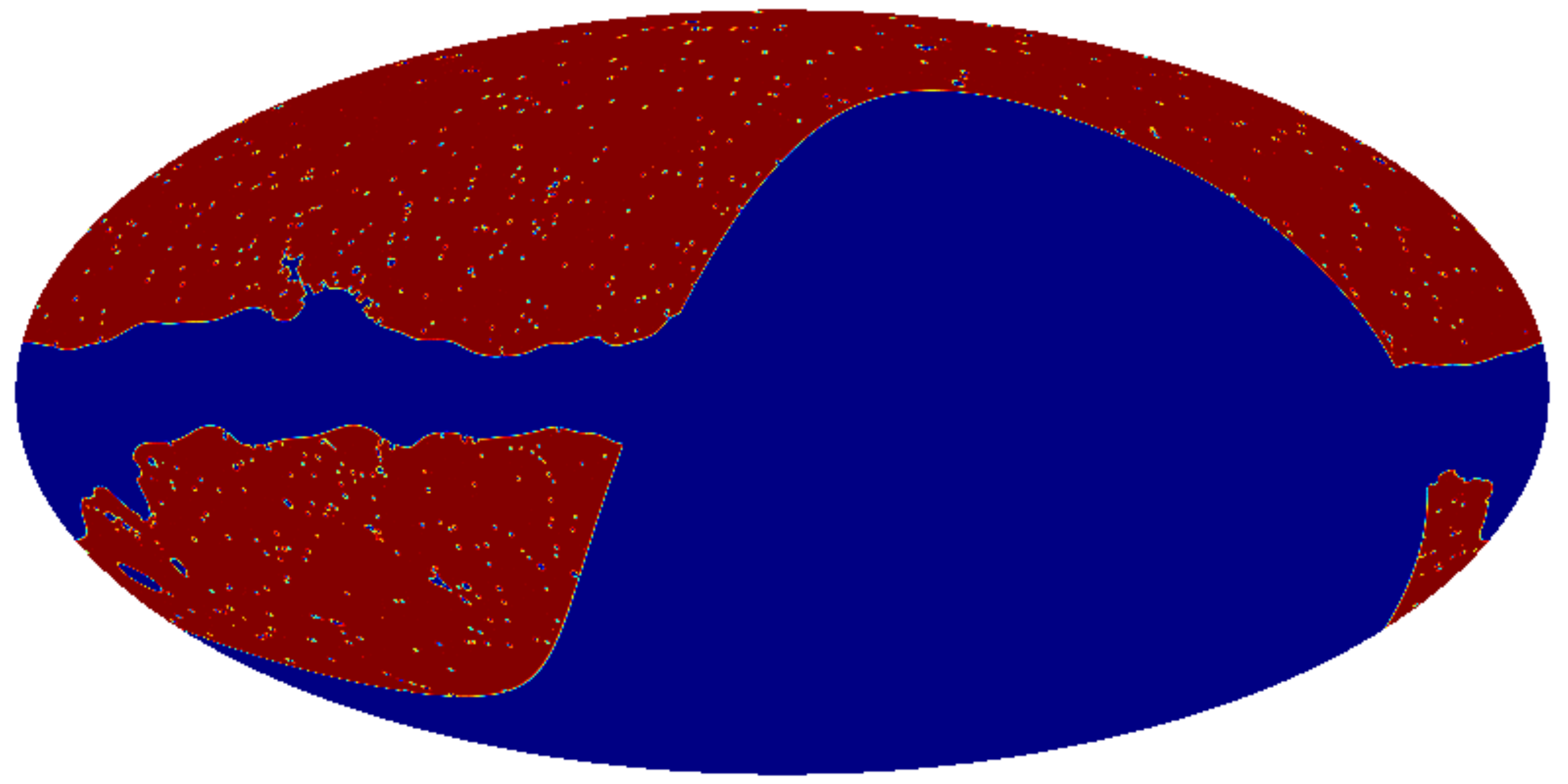}
\caption{\label{masks_fig} Apodized masks for the southern (\emph{left}) and northern (\emph{right}) ecliptic hemispheres.
}
\end{figure}

\begin{figure}
\centering
\includegraphics[width=13.5cm]{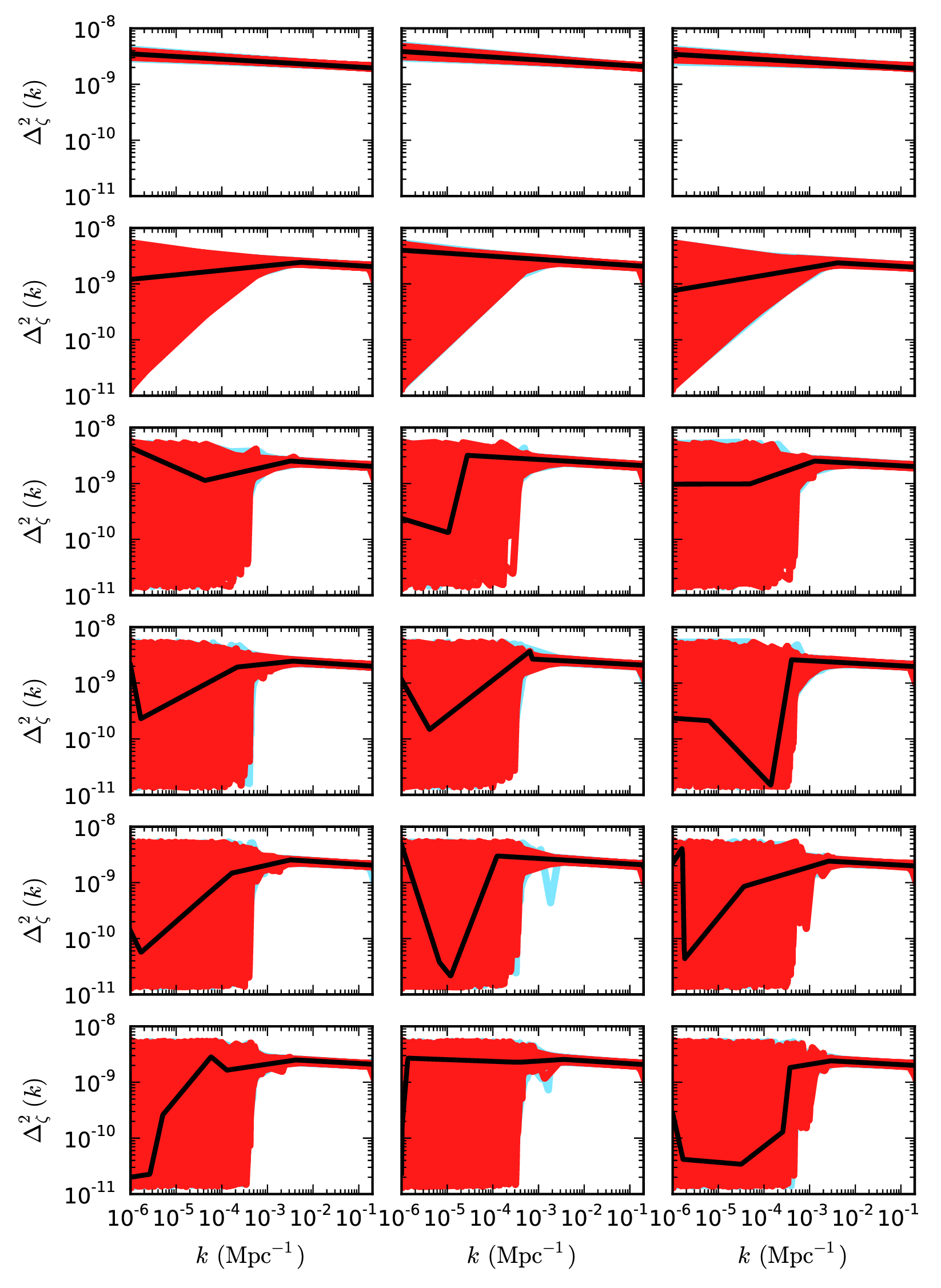}
\caption{\label{ps_all_hemispheres_lin_fig} The reconstructed primordial power spectrum (PPS) on the full sky compared to different ecliptic hemispheres.  The columns correspond to (\emph{left}) the full sky; (\emph{middle}) the southern hemisphere; and (\emph{right}) the northern hemisphere. The number of knots increases from $0$ to $5$ from top to bottom. The black solid lines show the best-fit PPS, the red lines are the PPS in the $68\%$ CI, and the light blue lines are the PPS in the $95\%$ CI.  All of the plots have been obtained from the SMICA map from \emph{Planck} using the linear-spline interpolation model with $n$ knots.
}
\end{figure}

\begin{figure}
\centering
\includegraphics[width=10cm]{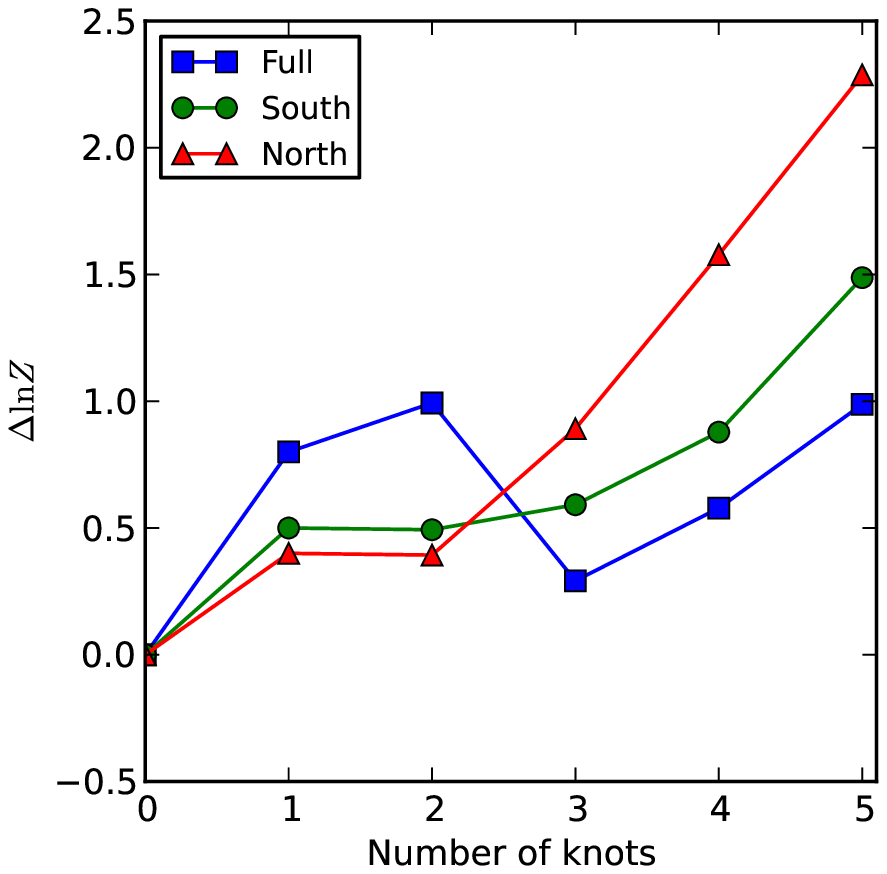}
\caption{\label{hemispheres_lin_fig} Change in Bayesian evidence $\Delta \ln Z$ with respect to $\Lambda$CDM as a function of the number of knots used in the primordial power spectrum reconstruction. The blue squares correspond to the full sky, the green circles correspond to the southern hemisphere (in ecliptic coordinates), and the red triangles correspond to the northern hemisphere. All of the plots have been obtained from the SMICA map from \emph{Planck} using the linear-spline interpolation model with $n$ knots.
}
\end{figure}

The latest results from \emph{Planck} \cite{Collaboration:2013vj} confirm the presence \cite{Eriksen:2003db,2009ApJ...704.1448H} of a power asymmetry in different ecliptic hemispheres of the sky.  Only about $4$ simulations out of $500$ have a higher level of power asymmetry than that found in the data \cite{Collaboration:2013vj}.  While some authors (see \emph{e.g.}, Ref.~\cite{2013JCAP...09..033F}) have questioned the significance of the power asymmetry on small scales, it is a persistent anomaly on large scales.  There are many suggestions \cite{Erickcek:2008sm,Dai:2013kfa,Liddle:2013czu,McDonald:2014uj,Rath:2014td,Aslanyan:2013jwa} as to how such an asymmetry might arise, and we study the possibility of the power asymmetry resulting from having more structure in the primordial power spectrum (PPS) in one hemisphere compared to the other.

Using the SMICA map from Planck \cite{Collaboration:2013vx} we reconstruct the PPS in each hemisphere, including the mask as described in Section~\ref{sec_simulations}. The likelihood  is computed using the same process employed with the simulated maps in Section~\ref{sec_simulations}. The apodized masks for the two hemispheres used in our analysis are shown in Fig.~\ref{masks_fig} and we test our mask and statistical methods by reproducing the \emph{Planck} power spectra on different hemispheres of the SMICA map (cf. Fig.~28 in Ref.~\cite{Collaboration:2013vj}). We also made a full analysis of the SMICA map with our combined mask and a standard power-law PPS as in Eq.~\eqref{pps_standard}. The resulting constraints on the cosmological parameters are very similar to those obtained by \emph{Planck} \cite{Collaboration:2013vx} (see Ref.~\cite{Aslanyan:2013ts} for a detailed comparison).

We reconstruct the PPS on the full sky, as well as the northern and the southern ecliptic hemispheres, using linear-spline interpolation ($\mathrm{LS}_n$) with $n \le 5$ knots. The resulting posterior distributions for the PPS are shown in Fig.~\ref{ps_all_hemispheres_lin_fig}, and the Bayesian evidence ratios as a function of the number of knots are shown in Fig.~\ref{hemispheres_lin_fig}. Although small differences are seen in the posterior distributions of the PPS and the best-fit PPS between the two hemispheres,  Bayesian evidence does not give significant preference to extra features in either hemisphere. The  evidence ratios for the two hemispheres are very close to each other, and to the full sky case.  The largest evidence value seen is $\Delta \log Z =2.3$ for the northern hemisphere but this was found with the foreground cleaned SMICA map.  Given that the fixing of the foreground parameters for Planck led to a spurious increase in evidence on the full sky, we interpret these results as  indicating that there is no difference in the structure of the PPS in the two hemispheres.

\section{Summary}\label{sec_summary}

We have applied the ``knot-spline'' reconstruction method \cite{2012JCAP...06..006V} to the Planck temperature data. This paper breaks new ground by checking the algorithm's ability to recover the PPS from simulated maps with artificially introduced features and by confirming that cosmological parameter constraints obtained from the Planck data are not diluted  when  the usual assumptions about the form of the primordial power spectrum are relaxed. Furthermore, we  investigate whether the hemispherical power asymmetry visible in the WMAP and Planck temperature maps is correlated with differences in the primordial power spectrum recovered from each hemisphere, finding that the two power spectra are in good agreement. Finally, the numerical tools needed to reproduce or extend this analysis are now included with the \textsc{Cosmo++} library \cite{Aslanyan:2013ts}.

The PPS reconstruction method used here allows the location of the knots to vary in both $k$-space and amplitude, and allows us to capture both gentle variations in the spectrum as well as a broad class of localized features.  Increasing the possible complexity of the PPS necessarily improves the fit to the data, and we must guard against the ``look-elsewhere'' effect or ``fitting the noise''. Determining the optimal number of knots can be posed as a model selection problem, and we use Bayesian evidence ratios to safeguard against overfitting.

We applied our methods to simulated maps with Planck characteristics to check the reliability of the method, and to estimate the amplitude of possible features in the PPS that the method can detect. We were able to recover modulations which modified an underlying power law spectrum by less than $5\%$. Typically, specific modulations have well-defined thresholds above which they are very easy to detect; for example, a long wavelength modulation with an amplitude a factor of three beyond the threshold of detectability yields an improvement in evidence of $\Delta\ln Z=51$.  Because the \emph{Planck} likelihoods are more sensitive to features at $k>10^{-3} \, \impc$, the  posteriors on the knots' positions (Fig.~\ref{ps_sim_lin_fig}) show a decrement of power at $k<10^{-3} \, \impc$, although this is not a feature of the simulated data, and cautions us against over-interpreting an apparent decrement in large scale power in the actual sky maps.  More generally, the weak evidence computed for the smaller modulation is partly driven by the use of ``uninformative'' priors for the modulated spectrum \cite{Easther:2011yq}. Consequently, Bayesian evidence does not permit a strictly algorithmic solution to cosmological model-selection problems and with maximum entropy priors similar to those used here, and nuanced physical analyses of the improvement in the maximum likelihood along with cross-checks against other datasets will remain important.

Having tested our methods, we reconstructed the PPS from \emph{Planck} CMB temperature and lensing data.  We found no evidence for deviations from the standard power law PPS on  scales with $k \, \gtrsim \, 10^{-3} \, \impc$.  Although on larger scales the data is not able to distinguish between models with or without features due to cosmic variance, the extensions to $\Lambda$CDM do not have sufficient Bayesian evidence to favor them over a standard power-law PPS.  Furthermore, the posteriors for the ``standard''  cosmological parameters did not differ substantially from the power-law case did not change significantly  when a more general PPS was allowed and we can conclude that the \emph{Planck} constraints on these parameters are  robust.  Finally, we performed a PPS reconstruction on each individual hemisphere, but found no systematic difference between the results showing that any ``hemispherical anomaly'' is not associated with differences in the underlying power spectrum.

This paper is the first in a sequence of analyses of non-standard power spectra. In particular we will investigate the implications of the recent detection of $B$-mode polarization by the BICEP2 telescope \cite{Ade:2014xna,Ade:2014gua} for the scalar power spectrum  \cite{Abazajian:2014tqa}, and in a third paper we will study whether permitting a non--power-law PPS changes the estimated values of derived parameters such as $\sigma_8$ or modifies estimated constraints on the neutrino sector.

\acknowledgments

We thank Brendon Brewer for useful discussions.  The authors wish to acknowledge the contribution of the NeSI high-performance computing facilities and the staff at the Centre for eResearch at the University of Auckland. New Zealand's national facilities are provided by the New Zealand eScience Infrastructure (NeSI) and funded jointly by NeSI's collaborator institutions and through the Ministry of Business, Innovation and Employment's Infrastructure programme [\url{http://www.nesi.org.nz}].  GA acknowledges the use of Windows Azure Cloud Computing Services through the Windows Azure Research grant. KNA is supported by NSF CAREER Grant No. PHY-11-59224.

\appendix
\section{Numerical data}
\label{appendix}

In this appendix we present tables of numerical values to further quantify our analysis.  Table~\ref{knot_bounds_table}  reports the numerical values for the posterior probabilities on the location of knots in the reconstructed primordial power spectrum in Figs~\ref{pps_planck_fig}~and~\ref{knot_bounds_fig}.  Table~\ref{cosmo_params_table} shows the constraints on the cosmological parameters in Fig.~\ref{cosmo_params_fig}.

\begin{table}[p]
  \small
\renewcommand{\arraystretch}{1.5}
\setlength{\arraycolsep}{5pt}
\begin{eqnarray*}
\begin{array}{c||cc|cc|cc}
  & \multicolumn{2}{c}{\mathrm{LS}_n \mathrm{\, var.}} & \multicolumn{2}{c}{\mathrm{LS}_n \mathrm{\, fixed}} & \multicolumn{2}{c}{\mathrm{CS}_n \mathrm{\, fixed}} \\
 n & \ln k_i & A_i & \ln k_i& A_i & \ln k_i & A_i \\
\hline
0 & -13.8 & 3.493_{-0.067}^{+0.064} & -13.8 & 3.466_{-0.056}^{+0.054} & -13.8 & 3.466_{-0.056}^{+0.054} \\
   & 0.0 & 2.967_{-0.073}^{+0.078} & 0.0 & 2.987_{-0.064}^{+0.067} & 0.0 & 2.987_{-0.064}^{+0.067} \\
\hline
1 & -13.8 & 1.2_{-2.1}^{+1.8} & -13.8 & 1.8_{-2.5}^{+1.7} & -13.8 & 2.91_{-0.56}^{+0.49} \\
   & -10.2_{-2.4}^{+2.9} & 3.35_{-0.11}^{+0.11} & -9.7_{-2.8}^{+8.1} & 3.32_{-0.23}^{+0.11} & -7.5_{-2.5}^{+2.8} & 3.157_{-0.146}^{+0.077} \\
   & 0.0 & 2.970_{-0.071}^{+0.074} & 0.0 & 2.981_{-0.166}^{+0.097} & 0.0 & 3.000_{-0.059}^{+0.065} \\
\hline
2 & -13.8 & 1.0_{-1.9}^{+2.0} & -13.8 & 1.1_{-2.0}^{+1.9} & -13.8 & 1.1_{-1.8}^{+1.6} \\
   & -11.7_{-1.4}^{+1.9} & 1.3_{-2.0}^{+1.7} & -11.4_{-1.6}^{+2.3} & 2.42_{-2.81}^{+0.97} & -10.0_{-1.8}^{+2.0} & 2.84_{-0.65}^{+0.31} \\
   & -8.5_{-2.3}^{+1.6} & 3.298_{-0.069}^{+0.089} & -7.8_{-2.6}^{+6.8} & 3.253_{-0.233}^{+0.099} & -5.1_{-2.6}^{+2.1} & 3.153_{-0.085}^{+0.064} \\
   & 0.0 & 2.974_{-0.071}^{+0.076} & 0.0 & 2.97_{-1.62}^{+0.11} & 0.0 & 3.004_{-0.061}^{+0.063} \\
\hline
3 & -13.8 & 1.1_{-2.1}^{+2.0} & -13.8 & 1.0_{-1.9}^{+2.0} & -13.8 & 1.1_{-1.9}^{+1.8} \\
   & -12.2_{-1.0}^{+1.6} & 1.0_{-2.0}^{+1.9} & -12.1_{-1.1}^{+1.8} & 1.2_{-2.1}^{+1.8} & -12.07_{-0.84}^{+1.16} & 1.3_{-1.8}^{+1.4} \\
   & -10.3_{-1.7}^{+1.5} & 1.5_{-2.3}^{+1.5} & -9.8_{-1.9}^{+1.8} & 3.07_{-3.00}^{+0.29} & -9.4_{-1.4}^{+1.5} & 2.75_{-0.90}^{+0.41} \\
   & -8.1_{-1.7}^{+1.4} & 3.278_{-0.060}^{+0.069} & -7.0_{-2.2}^{+6.3} & 3.219_{-0.219}^{+0.094} & -5.6_{-1.7}^{+1.9} & 3.168_{-0.068}^{+0.056} \\
   & 0.0 & 2.976_{-0.068}^{+0.073} & 0.0 & 2.95_{-2.60}^{+0.12} & 0.0 & 2.999_{-0.061}^{+0.055} \\
\hline
4 & -13.8 & 1.0_{-1.9}^{+1.9} & -13.8 & 1.1_{-2.0}^{+1.9} & -13.8 & 0.9_{-1.8}^{+1.8} \\
   & -12.62_{-0.73}^{+1.26} & 0.9_{-1.9}^{+2.0} & -12.47_{-0.87}^{+1.34} & 1.1_{-2.0}^{+1.9} & -12.62_{-0.64}^{+0.96} & 1.0_{-1.7}^{+1.8} \\
   & -11.2_{-1.3}^{+1.4} & 1.0_{-1.9}^{+2.0} & -10.9_{-1.4}^{+1.6} & 1.3_{-2.1}^{+1.7} & -10.9_{-1.1}^{+1.2} & 1.6_{-1.9}^{+1.2} \\
   & -9.7_{-1.4}^{+1.2} & 1.7_{-2.3}^{+1.4} & -9.0_{-1.8}^{+1.4} & 3.19_{-2.56}^{+0.16} & -8.9_{-1.2}^{+1.4} & 2.80_{-0.95}^{+0.37} \\
   & -8.0_{-1.3}^{+1.5} & 3.274_{-0.060}^{+0.059} & -5.0_{-3.3}^{+4.4} & 3.14_{-0.15}^{+0.15} & -5.6_{-1.5}^{+2.0} & 3.170_{-0.073}^{+0.052} \\
   & 0.0 & 2.972_{-0.073}^{+0.075} & 0.0 & 2.91_{-2.94}^{+0.16} & 0.0 & 3.003_{-0.058}^{+0.063} \\
\hline
5 & -13.8 & 1.0_{-1.9}^{+1.9} & -13.8 & 1.0_{-2.0}^{+1.9} &  &  \\
   & -12.53_{-0.78}^{+1.27} & 1.0_{-1.9}^{+2.0} & -12.65_{-0.74}^{+1.29} & 1.0_{-2.0}^{+1.9} & & \\
   & -11.1_{-1.3}^{+1.4} & 1.0_{-1.9}^{+1.9} & -11.4_{-1.2}^{+1.4} & 1.0_{-2.1}^{+1.9} &  & \\
   & -9.6_{-1.4}^{+1.2} & 1.7_{-2.4}^{+1.4} & -10.0_{-1.5}^{+1.3} & 1.4_{-2.2}^{+1.7} & & \\
   & -8.0_{-1.2}^{+1.5} & 3.274_{-0.056}^{+0.057} & -8.5_{-1.6}^{+1.2} & 3.22_{-1.96}^{+0.14} & & \\
   & -0.84_{-0.58}^{+0.52} & 3.010_{-0.066}^{+0.075} & -1.54_{-6.65}^{+1.00} & 3.10_{-0.11}^{+0.18} & & \\
   & 0.0 & 1.1_{-2.0}^{+1.9} & 0.0 & 2.90_{-3.08}^{+0.18} & & \\
\hline
\end{array}
\end{eqnarray*}
\caption{Numerical values for the constraints on the location of knots.  Companion to Figs~\ref{pps_planck_fig}~and~\ref{knot_bounds_fig}.}
\label{knot_bounds_table}
\end{table}

\begin{table}
  \small
\renewcommand{\arraystretch}{1.5}
\setlength{\arraycolsep}{5pt}
\begin{eqnarray*}
\begin{array}{c|cccc}
n & \Omega_bh^2 & \Omega_ch^2 & h & \tau \\
\hline
   & 0.02218_{-0.00032}^{+0.00031} & 0.1186_{-0.0030}^{+0.0029} & 0.684_{-0.013}^{+0.015} & 0.087_{-0.029}^{+0.031} \\
0 & 0.02225_{-0.00025}^{+0.00027} & 0.1182_{-0.0029}^{+0.0027} & 0.686_{-0.012}^{+0.013} & 0.092_{-0.027}^{+0.028} \\
   & 0.02225_{-0.00025}^{+0.00027} & 0.1182_{-0.0029}^{+0.0027} & 0.686_{-0.012}^{+0.013} & 0.092_{-0.027}^{+0.028} \\
\hline
   & 0.02217_{-0.00030}^{+0.00031} & 0.1186_{-0.0029}^{+0.0028} & 0.684_{-0.013}^{+0.014} & 0.089_{-0.029}^{+0.030} \\
1 & 0.02226_{-0.00025}^{+0.00025} & 0.1182_{-0.0028}^{+0.0029} & 0.686_{-0.013}^{+0.013} & 0.091_{-0.026}^{+0.029} \\
   & 0.02241_{-0.00026}^{+0.00029} & 0.1173_{-0.0026}^{+0.0026} & 0.691_{-0.012}^{+0.013} & 0.107_{-0.027}^{+0.029} \\
\hline
   & 0.02220_{-0.00030}^{+0.00031} & 0.1184_{-0.0028}^{+0.0028} & 0.685_{-0.013}^{+0.014} & 0.091_{-0.030}^{+0.029} \\
2 & 0.02226_{-0.00024}^{+0.00023} & 0.1181_{-0.0026}^{+0.0025} & 0.687_{-0.012}^{+0.012} & 0.094_{-0.026}^{+0.027} \\
   & 0.02233_{-0.00033}^{+0.00032} & 0.1179_{-0.0028}^{+0.0028} & 0.688_{-0.013}^{+0.014} & 0.101_{-0.029}^{+0.033} \\
\hline
   & 0.02220_{-0.00030}^{+0.00031} & 0.1184_{-0.0028}^{+0.0028} & 0.685_{-0.013}^{+0.014} & 0.091_{-0.028}^{+0.029} \\
3 & 0.02225_{-0.00024}^{+0.00023} & 0.1183_{-0.0028}^{+0.0025} & 0.686_{-0.011}^{+0.012} & 0.093_{-0.026}^{+0.027} \\
   & 0.02230_{-0.00032}^{+0.00029} & 0.1180_{-0.0025}^{+0.0027} & 0.688_{-0.013}^{+0.012} & 0.101_{-0.028}^{+0.026} \\
\hline
   & 0.02219_{-0.00030}^{+0.00031} & 0.1186_{-0.0029}^{+0.0029} & 0.684_{-0.013}^{+0.014} & 0.090_{-0.029}^{+0.030} \\
4 & 0.02225_{-0.00024}^{+0.00024} & 0.1181_{-0.0025}^{+0.0026} & 0.687_{-0.012}^{+0.011} & 0.094_{-0.027}^{+0.028} \\
   & 0.02232_{-0.00030}^{+0.00029} & 0.1178_{-0.0027}^{+0.0029} & 0.688_{-0.013}^{+0.013} & 0.102_{-0.029}^{+0.030} \\
\hline
   & 0.02220_{-0.00030}^{+0.00032} & 0.1184_{-0.0029}^{+0.0028} & 0.685_{-0.013}^{+0.014} & 0.092_{-0.027}^{+0.031} \\
5 & 0.02225_{-0.00024}^{+0.00026} & 0.1180_{-0.0028}^{+0.0027} & 0.687_{-0.012}^{+0.013} & 0.096_{-0.026}^{+0.027} \\
\hline
\end{array}
\end{eqnarray*}
\caption{Cosmological parameter constraints as a function of the number of knots $n$ at the $68\%$ CI. For each value of $n$ the first line corresponds to linear-spline interpolation ($\mathrm{LS}_n$) with varying foreground paramters, the second line is $\mathrm{LS}_n$ with fixed foreground parameters, and the third line is cubic-spline interpolation ($\mathrm{CS}_n$) with fixed foreground parameters. Companion to Fig.~\ref{cosmo_params_fig}.}
\label{cosmo_params_table}
\end{table}

\bibliography{papers_lib,citations,kevork}

\end{document}